# On the Performance comparison of RIP, OSPF, IS-IS and EIGRP routing protocols

Vasos Hadjioannou
Department of Computer Science
University of Nicosia
Nicosia, Cyprus
e-mail: hadjioannou.v@student.unic.ac.cy

*Abstract*— **Nowadays, routing protocols have become a crucial part of the modern communication networks. A routing protocol's responsibility lies in determining the way routers communicate with each other in order to forward any kind of packets, from a source to a destination, using the optimal path that would provide the most efficiency. There are many routing protocols out there today, some old and some new, but all are used for the same purpose. In general, to ideally select routes between any two nodes on a computer network and disseminate information. This paper takes into consideration four of such routing protocols (RIP, OSPF, IS-IS and EIGRP), expresses them and analyzes their way of operation. It also presents the results of a simulation, that took place for the sole purpose of studying the behavior of those four protocols, under the same circumstances, as well as the evaluation of the comparison with one another based on the results of the simulation.**

## I. INTRODUCTION

When talking about routing protocols, a very important question inevitably comes up: "Which protocol to use?". The full-proof way to use in order to decide, is to set up a network and use the protocols that might possibly suit the network's needs. This will prove to be extremely inefficient and expensive in time, money and resources. To give an answer to this question, network simulations come into play.

A network simulation is a technique for modeling the behavior of a network by calculating the interaction between the different network entities, using mathematical formulas, or actually capturing and playing back observations from a production network. Such simulations can be performed by various tools (such as OPNET, ns, NetSim etc.) otherwise known as network simulators.

A network simulator is a software that was designed to predict the behavior of a computer network by evaluating the performance of a model that corresponds to the network that needs to be analyzed.

In this case, the OPNET simulator was used to evaluate the performance of the four aforementioned protocols and compare them with each other, by sending IP datagrams and traffic from one computer to another, over a modeled network.

## II. DISTANCE VECTOR ROUTING PROTOCOLS

### A. Routing Information Protocol (RIP)

RIP is a distance-vector routing protocol that is based on the Bellman-Ford algorithm and uses hops as a routing metric. It avoids loops by limiting the number of hops that are allowed in a single path, from a source to a destination. The maximum number of hops allowed are 15, which limits the size of networks RIP can support. If a route contains more than 15 hops, it is then considered as an infinite distance and its destination is unreachable. RIP uses the User Datagram Protocol (UDP) as its transport protocol, and is assigned the reserved port number 520.

Bellman-Ford algorithm:

The Bellman–Ford algorithm is an algorithm that computes shortest paths from a single source vertex to all of the other vertices in a weighted graph. It is similar to Dijkstra, but instead of greedily selecting the minimum-weighted node, not yet processed, to relax, it simply relaxes all edges, and does this |v|-1 times. The repetitions allow minimum distances to propagate accurately, since in the absence of negative cycles, the shortest path can visit each node at most once.

Bellman-Ford can detect cycles and report their existence, and its running time is O(|v|*|e|) where *v* denotes the number of vertices in the directed graph, and *e* the number of edges [1].

```
BELLMAN-FORD(G, w, s)
1  INITIALIZE-SINGLE-SOURCE(G, s)
2  for i ← 1 to |V[G]| − 1
3      do for each edge (u, v) ∈ E[G]
4          do RELAX(u, v, w)
5  for each edge (u, v) ∈ E[G]
6      do if d[v] > d[u] + w(u, v)
7          then return FALSE
8  return TRUE
```

Figure 1: Bellman-Ford algorithm pseudo code

RIP operation:
Every RIP router maintains a routing table, and for each destination there is one routing entry:

- the shortest distance from the router to the destination
- next router along the path to the destination

The distance from the router to the destination is periodically advertised by the routers to their neighbors. When a router receives such an advertisement, it checks whether or not any of the advertised routes are useful in improving the current routes. If so, the current route of the router is updated, in order for it to go through the advertising neighbor.

As mentioned before, routes are compared by their length, where their value needs to be between 1 and 16 (16 means infinity). In order for loops to be avoided, routers must do split horizon when they advertise a route to their neighbor.

The complete routing table is advertised by a router to all of its neighbors, and for each route there are two timers ("timeout" and "garbage-collection" time).

A destination will be marked as unreachable when a route hasn't been refreshed for three minutes, letting the router assume that there is a link failure. This will set the timer of the garbage collection to 2 minutes, and if the timer expires before the entry is updated, the route will then be removed from the routing table [2].

RIP disadvantages:
- Maximum hop size is 15
- Variable length subnet masks are not supported by RIP version 1
- Slow convergence and count-to-infinity problem

*B. Enhanced Interior Gateway Routing Protocol(EIGRP)*

The Enhanced Interior Gateway Routing Protocol is an advanced distance-vector routing protocol, designed by Cisco Systems, in ordered to help automate routing decisions and configuration on a computer network. EIGRP was designed to support classless IPv4 addresses, and for this reason, it replaced the Interior Gateway Routing Protocol (IGRP), which could not.

Routers use EIGRPs discovery and recovery mechanism in order to dynamically learn of other routers on their directly attached networks. They must also find out whether or not their neighbors are reachable, or indeed operational, which is achieved by having each router send small Hello packets periodically. A router can determine the operation of its neighbor as long as Hello packets keep getting received by it, and once the availability of its neighbors is made positive, routers can exchange routing information [3].

The rules by which traffic is forwarded is kept in a routing table which many routers have, and if a valid path to a destination is not known by a router, the traffic is discarded. Furthermore, routers automatically share information, which makes EIGRP a dynamic routing protocol.

EIGRP is an enhanced distance-vector protocol which relies on the Diffused Update Algorithm (DUAL) to calculate the shortest path to a network.

EIGRP has four methods[4]:
- Neighbor Discovery/Recovery
- Reliable Transport Protocol (RTP)
- Diffusion Update Algorithm (DUAL)
- Protocol Dependent Modules (PDM)

III. INTERIOR GATEWAY PROTOCOLS

Both of the following Interior Gateway Algorithms (OSPF and IS-IS) are based on the Dijkstra algorithm.

Dijkstra:

The Dijkstra algorithm is a graph search algorithm that solved the single-source shortest path problem for a graph with non-negative edge paths, producing a shortest path tree. It maintains a set of vertices (S) whose shortest path from the source has already been determined. It repeatedly selects the vertex with the minimum shortest path, adds it to the set, and then relaxes all edges leaving from it.

Steps for executing the Dijkstra algorithm:
1. Assign to every node distance value from the source
   - 0 for the source
   - ∞ for all other nodes
2. Mark all nodes unvisited and set the initial node as "current". Create a set of unvisited nodes consisting all the nodes except the initial one.
3. For the current node, consider all of its unvisited neighbors and calculate their tentative distances. If the new distance is less than what already exists, we override the old distance with the new one.
4. When we are done considering all the neighbors of the current node, mark it as "visited" and remove it from the "unvisited" set.
5. If the destination node has been marked "visited" ---> exit the search
6. Otherwise, select the unvisited node with the smallest tentative distance and mark it as *current*. Repeat from step3.

*A. Open Short Path First (OSPF)*

OSPF is an Interior Gateway Protocol (IGP), based on the Dijkstra algorithm, used for routing Internet Protocol (IP) packets within autonomous systems, or other routing domains. It is a link-state routing protocol. Such protocols are also referred as SPF (Shortest Path First)-based or distributed database protocols [5].

In link-state routing, each node constructs a map, in the form of a graph, which contains the connectivity information about the network. This map shows how the nodes are connected to one another and which node to which. Every node will then calculate the best logical path to every possible destination in the network, leading to the creation of the node's routing table, by collecting the best paths individually calculated by each node. After the routing table is created, it is presented to the Internet Layer which routes datagrams according to the address located in the IP packets [6].

OSPF is able to detect link failures, or any other changes in the topology, and adapt within seconds by converging to a new loop-free structure. Shortest Path First algorithms, in general, make for a stable network since they require frequent, but small, updates and converge quickly, thus avoiding routing loops and the count-to-infinity problem. On the other hand, they take up a lot of CPU power and memory.

OSPF supports Internet Protocol Version 4 (IPv4), Internet Protocol Version 6 (IPv6), Variable Subnet Masking (VLSM) and Classless Inter-Domain Routing (CIDR) addressing models.

The portable devices (e.g. smart phones, tablets etc.) in emerging mobile network architectures have shrunk in size, incorporating advanced functions and mechanisms as explored in [7-15]. Authors in [16] use a novel algorithm to handle the traffic with the self-similarity properties that the traffic that traverses the routing paths exhibits. In [17-19] the traffic diversities are used incorporated using a relay routing mechanism for MP2P systems. Authors in [20-21] use a resource traffic offloading technique for energy usage optimization in the cloud, using the centrality principle of social networks. Mobile users take advantage of the mobile opportunistic cloud, in order to increase their reliability in service provision by guaranteeing sufficient resources for the execution of mobile applications. In [22] authors apply mechanisms for precise performance analysis of reactive routing protocols in Mobile Ad hoc Networks including comparative evaluation of the efficiency under sudden failures. Very important comparative evaluations have been exposed in [24-28] in both Cognitive radio level architecture and heuristic routing decisions. In this work a comparison of four diversified routing decisions are presented exposing both the pros and cons of each one the decisions/ implementations.

*B. Intermediate System to Intermediate System (IS-IS)*

IS-IS is a dynamic routing protocol designed and used for forwarding traffic within a packet-switched network by determining the best route for datagrams to take. It is also based on the Dijkstra algorithm, and similarly to OSPF, IS-IS is also an Interior Gateway Protocol (IGP), which uses link-state routing by reliably flooding link state information throughout a network of routers. Each Intermediate System (IS or router), that is configured to run IS-IS, creates a database of the network's topology aggregating the information[23].

Features of IS-IS [29]:
- Hierarchical routing
- Classless behavior
- Rapid flooding of new information
- Fast convergence
- Scalable

Operation of IS-IS:
Each router running IS-IS sends a hello packet to every interface that has IS-IS enabled in order to discover its neighbors and establish adjacencies. If the information contained in a hello packet meets the criteria for forming an adjacency, then routers that share a common data link will become IS-IS neighbors. After that, routers flood Link-State Packets (LSP) to their neighbors, except the ones who they received the same LSP from. Then, all routers will construct their individual Link-State Database and a shortest-path tree (SPT) is calculated by each IS, which is used to construct the routing table [8].

IV. SIMULATION

The simulation was performed using Opnet IT Guru Academic Edition [30] and the model was setup with the help of many video and written tutorials. A network was created, consisting of 2 computers that exchange IP packets with one another through several routers that are in between them.

A failure scenario also exists where the communication between 2 routers would repeatedly fail and recover, in order to investigate the behavior of the routing protocols in such cases. Furthermore, the route each protocol takes will be investigated by replacing the links of a particular route with faster ones, in order to check whether the datagram packets will travel from the fastest or the shortest route to reach their destination.

*A. Network model*

The created network is comprised of the following components:
- Ethernet workstation: Workstation used with client-server applications that runs over TCP/IP and UDP/IP. It can support one underlying Ethernet connection at 10Mbps, 100Mbps, or 1000Mbps and for each packet, it requires a fixed amount of time in order to route them, as the "IP Forwarding Rate" attribute of the node specifies. Packets are routed on first-come-first-serve basis and, depending on the transmission rates of the corresponding output interfaces, they may encounter queuing at the lower protocol layers,.
- Ethernet64 switch: Switch that implements the Spanning Tree Algorithm in order for the network topology not to contain any loops. It supports up to 64 Ethernet interfaces. The communication between switches is done by sending Bridge Protocol Data

- Units (BPDUs) to each other, and the packets are received and processed based on the current configuration of the spanning tree.
- Cisco 7000 router: An IP-based router gateway model. According to the destination IP address, packets that arrive on an IP interface are routed to the appropriate output interface.
- 10BaseT duplex link: Ethernet connection link that transmits data at a rate of 10Mbps and is able to connect any Stations, Hubs, Bridges, Switches and LAN nodes with one another.
- 100BaseT duplex link: Similar to the 10BaseT duplex link. The only difference is that the speed of data transfer over the 100BaseT duplex link is 100Mbps instead of 10Mbps.

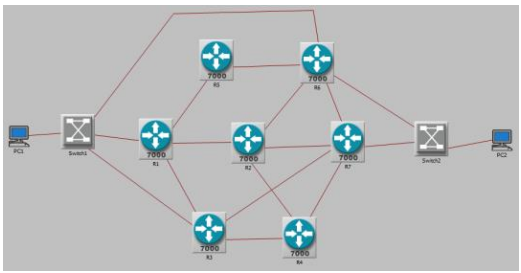

Figure 2: Topology of the network

The duration of the simulation is set to be 15 minutes with scheduled failures between the routers R1 and R2. The schedule of the failures is as follows:

| Time (seconds) | Status |
|---|---|
| 225 | Fail |
| 400 | Recover |
| 535 | Fail |
| 590 | Recover |
| 605 | Fail |
| 620 | Recover |
| 625 | Fail |
| 630 | Recover |
| 730 | Fail |
| 830 | Recover |

Table1: Timing of failures/recoveries

### B. Results

As mentioned above, the protocols that are compared are the Routing Information Protocol (RIP), Enhanced Interior Gateway Routing Protocol (EIGRP), Open Short Path First (OSPF) and Intermediate System to Intermediate System (IS-IS).

Traffic dropped: The number of IP datagrams dropped by all nodes in the network across all IP interfaces.

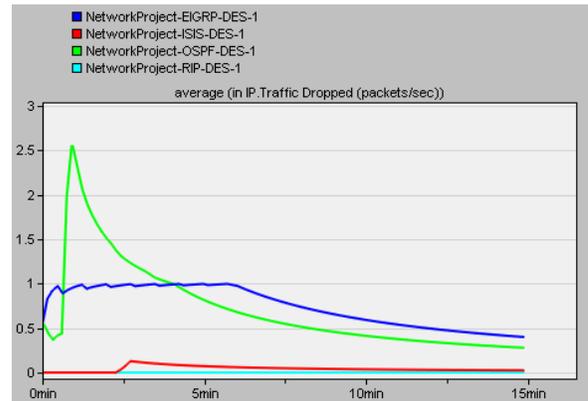

Figure 3: Average traffic dropped by the four protocols

As it is shown from Figure2, the protocol with the most traffic drops in case of a link failure is OSPF and then EIGRP. IS-IS doesn't drop many packages while RIP doesn't drop any.

Network convergence activity: Records square wave alternating between ordinates 0 and 1. While the there are no changes in the forwarding tables in the network, the value is 0. On the other hand, when some forwarding tables show signs of convergence, meaning there has been a change in the tables, the value becomes 1.

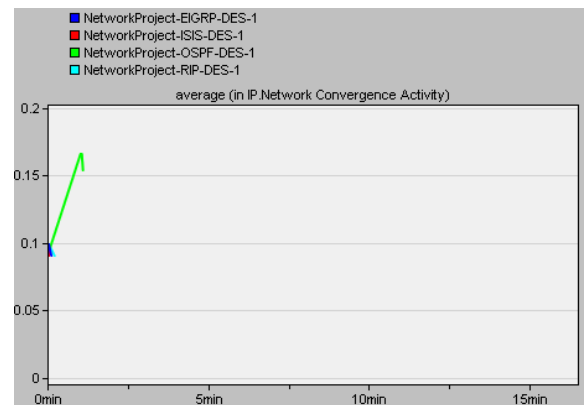

Figure 4: Average Network Convergence Activity

While EIGRP, IS-IS and RIP have almost the same network convergence activity, for even less than a minute, OSPF shows dramatic difference when compared to the other 3 protocols, in both magnitude and longevity.

Background Traffic: Measure of the end to end delay, experienced by information about background traffic flow while it travels between the flow's source and the flow's destination.

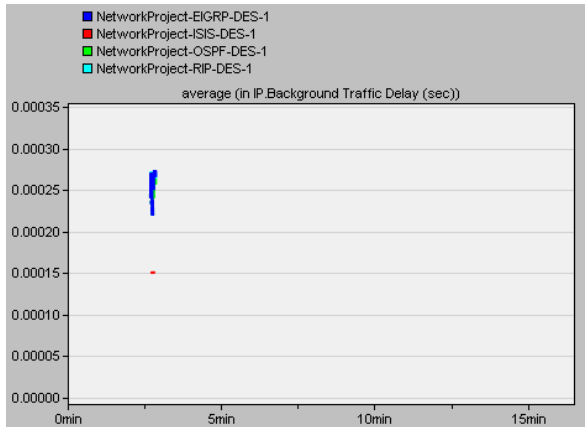

Figure 5: Average background traffic

In the case of background traffic, we can tell from Figure4 that IS-IS has the less traffic while the other 3 protocols have more or less the same.

To continue measuring, IP Unicast traffic was created with intensity of 100Packets/sec, 120000Bits/sec and 3600sec. of duration. The following measurements were taken (the results of the previous simulation did not change since it was simply an exchange of IP datagrams):

Delay: This statistic represents the end-to-end delay of all packets received by all the stations.

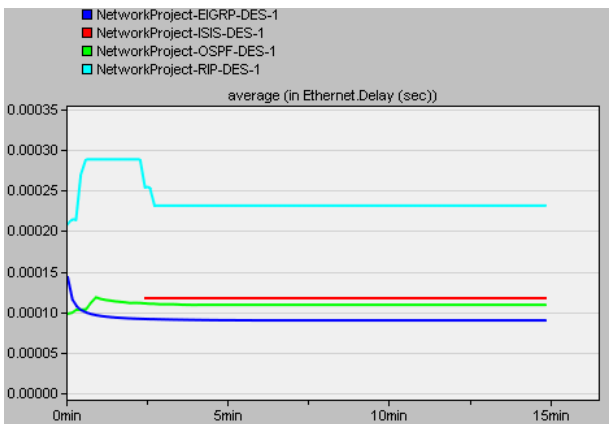

Figure 6: Average packet end-to-end delay

Figure 5 shows that the most delay was caused by RIP. While the other protocols had more or less the same amount of delay, EIGRP seems to be the fastest one.

Lastly, as mentioned before, some of the 10BaseT links were replaced with 100BasetT ones. As you can see from the network topology, the shortest path from PC1 to PC2 is no doubt PC1 -> Switch1 -> Router6 -> Switch2 -> PC2. For this reason, the highlighted links were replaced with faster links in order to see whether or not the protocols will keep using the same route, by investigating the number of hops from one PC to the other.

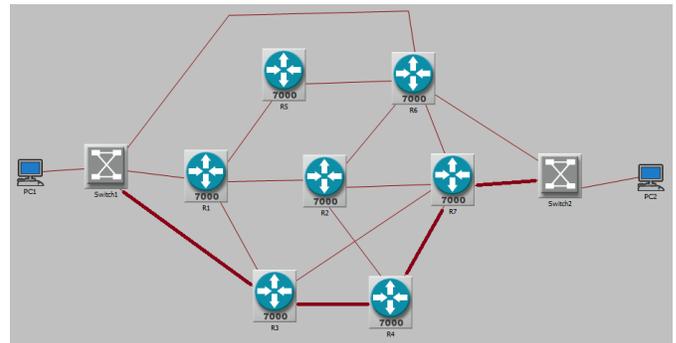

Figure 7: New topology with faster links on the highlighted part

Number of hops: This statistic gives the average number of IP hops taken by data packets reaching at a destination node.

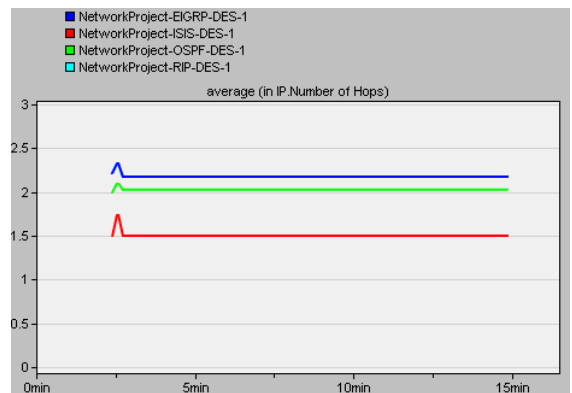

Figure 8: Average number of hops before the link replacement

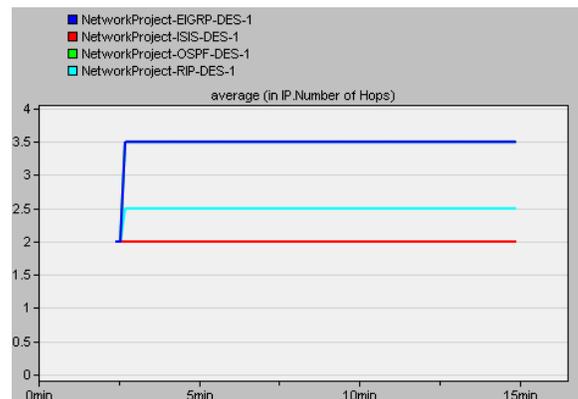

Figure 9: Average number of hops after the link replacement

We can see from the 2 figures that when all the links were 10BaseT, the average number of hops the different protocols would take was between 1.5 and 2.2

(Figure 1), which indeed indicates to the shortest path aforementioned. When the 10BaseT links were replaced with 100BaseT ones (Figure 2), we can see that the average number of hops for the protocols was increased to between 2 and 3.5 (OSPG and EIGRP overlap in both cases. That's why OSPF is not visible from the figures).

With this information we can deduct that all the protocols chose to follow the faster route even if it meant taking the more length one, rather than taking the shortest and slower route.

## V. CONCLUSION

After the series of measurements, we can see that in case of failure, the best protocols are RIP and IS-IS. Of course RIP is not suitable for large networks since it supports only 15 hops. It is indeed the best in the case of traffic drops but IS-IS is superior, even though its simulations were worse (but very close) to RIP, because of the fact that it supports larger networks.

When the average network convergence activity was measured, all the protocols showed more or less the same behavior, except from OSPF, which showed much more activity than the other three protocols.

For the average background traffic, IS-IS was the most efficient with the least background traffic, compared to the RIP, OSPF and EIGRP protocols that gave similar results to each other.

In regards of the average packet end-to-end delay, the protocol that did absolutely worse than the rest was the RIP protocol, and while the results of the other three were close, the one that showed more efficiency was EIGRP.

Finally, when it came to making a routing decision between the shortest and fastest path, all the protocols chose to follow the path that would transfer the datagrams faster from the source to the destination, in spite of the number of intermediate routers that they had to go through.